\documentclass[a4paper,10pt]{amsart}
\usepackage[utf8]{inputenc}
\usepackage{amsaddr}
\usepackage{calrsfs}
\usepackage{graphicx,color}
\usepackage[english]{babel}
\usepackage{graphicx}
\usepackage{amsmath}
\usepackage{epstopdf}
\usepackage{float}
\usepackage{color}

\usepackage{amssymb}

\numberwithin{equation}{section}

\theoremstyle{definition}

\theoremstyle{plain}
\newtheorem{thm}{Theorem}[section]

\theoremstyle{definition}
\newtheorem{rem}{Remark}[section]

\newcommand{\E}{\mathbb{E}}

\begin{document}
\title[Currency option pricing]
{Subdiffusive fractional Brownian motion regime for pricing currency options under transaction costs}

\date{\today}

\author[Shokrollahi]{Foad Shokrollahi}
\address{Department of Mathematics and Statistics, University of Vaasa, P.O. Box 700, FIN-65101 Vaasa, FINLAND}
\email{foad.shokrollahi@uva.fi}
\begin{abstract}
A new framework for pricing European currency option is developed in the case where the spot
exchange rate follows a subdiffusive fractional Brownian motion. An analytic formula for pricing European
 currency call option is proposed by a mean self-financing delta-hedging argument
in a discrete time setting. The minimal price of a currency option under transaction costs is obtained as time-step $\Delta t=\left(\frac{t^{\alpha-1}}{\Gamma(\alpha)}\right)^{-1}\left(\frac{2}{\pi}\right)^{\frac{1}{2H}}\left(\frac{k}{\sigma}\right)^{\frac{1}{H}}$
, which can be used as the actual price of an option. In addition, we also show that
time-step and long-range dependence have a significant impact on option pricing.
\end{abstract}


\keywords{Tubdiffusion process;
Currency option;
Transaction costs;
Inverse subordinator process}

\subjclass[2010]{91G20; 91G80; 60G22}

\maketitle

\section{Introduction}

The classical and still most popular model of option pricing is the Black–Scholes $(BS)$ \cite{black}. It is
assumed that the price of risky asset $V(t)$ is governed by a geometric Brownian motion, that is

\begin{eqnarray}
V(t)=V_0 e^{\mu t+\sigma B(t) },\qquad V(0)=V_0>0
\label{eq:1}
\end{eqnarray}

where $\mu,$ and $\sigma$ are fixed and $B(t)$ is the Brownian motion.

Empirical research show that the $BS$ model cannot capture many of the characteristic features of prices, such as: long-range dependence, heavy-tailed and skewed marginal distributions, the lack of scale invariance, periods of constant values, etc. In 1983, Garman and Kohlhagen $(G-K)$ \cite{garman} presented a modified version of the $BS$ model for pricing currency option. However, some scholars have argued that
option pricing with utilizing the $G-K$ model based on Brownian motion, cannot satisfactorily model for pricing currency option because currencies differ from stocks in financial markets. Hence, they have proposed some generalization of the $G-K$ model to capture the phenomena  from stock markets \cite{garman,ho}. To capture these non-normal behaviors, many
researchers have considered other distributions with fat tails such as the
Pareto-stable distribution and the Generalized Hyperbolic Distribution
among others. Moreover, self-similarity and long-range dependence
have become important concepts in analyzing the financial time series.
There is strong evidence that the stock return has little or no
autocorrelation. Since fractional Brownian motion $(FBM)$ has two important
properties called self-similarity and long-range dependence, it has the
ability to capture the typical tail behavior of stock prices or indexes.

The fractional Brownian motion $(FBM)$ model is an extension of the $BS$ model, which displays the long-range dependence observed in empirical data. The $FBM$ model is given by

\begin{eqnarray}
\widehat{V}(t)=\widehat{V}_0exp\{\mu t+\sigma \widehat{B}_H(t)\},  \qquad \widehat{V}_0>0,
\label{eq:3}
\end{eqnarray}
  where $B_H(t)$ is a $FBM$ with Hurst parameter $H\in[\frac{1}{2},1)$. It has been shown that the $FBM$ model admits arbitrage in a complete and frictionless market \cite{cheridito,wang1,sottinen2003arbitrage,shokrollahi3,xiao2010pricing}. Wang \cite{wang} resolved this contradiction by giving up the arbitrage argument and examining option replication in the presence of proportional transaction costs in discrete time setting \cite{mastinvsek}.

Magdziarz \cite{magdziarz} applied the subdiffusive mechanism of trapping events
to describe properly financial data exhibiting periods of constant values and introduced the
subdiffusive geometric Brownian motion

\begin{eqnarray}
V_\alpha(t)=V(T_{\alpha}(t)),
\label{eq:2}
\end{eqnarray}

as the model of asset prices exhibiting subdiffusive dynamics, where $V_\alpha(t)$ is a subordinated
process (for the notion of subordinated processes please refer to Refs. \cite{janicki,janicki1,piryatinska}), in which the
parent process $V(\tau)$ is the geometric Brownian motion  defined in (\ref{eq:1}) and $T_{\alpha}(t)$ is the inverse $\alpha$-stable subordinator
defined in the following way
\begin{eqnarray}
T_{\alpha}(t)=inf\{\tau>0 :Q_\alpha(\tau)>t\},\quad0<\alpha<1,
\label{eq:7}
\end{eqnarray}

$ Q_\alpha(t)$ is the $\alpha$-stable subordinator with Laplace transform: $\E\left(e^{-\eta  Q_{\alpha}(\tau)}\right)=e^{-\tau\eta^\alpha}$, $0<\alpha<1$, where $\E$ denotes the mathematical expectation. Assuming
that $T_{\alpha}(t)$ is independent of the Brownian motion $B(t)$. Moreover, he demonstrated that this model is free-arbitrage but is incomplete. In this regard, he presented a new formula for fair prices of European option with the corresponding subdiffusive $BS$ model. For additional information about more models that describe such characteristic behavior, you can see \cite{scalas,janczura2011,gu,shokrollahi1,guo2017option}.

In this study, in order to capture the long-range dependence of interest rates and to examine option replication in the presence of proportional transaction costs in a discrete time setting, we consider the problem of pricing currency option, where the spot
exchange rate is governed by a subdiffusive $FBM$ as follows
\begin{eqnarray}
&&S_t=\widehat{V}(T_{\alpha}(t))=S_0exp\{\mu T_{\alpha}(t)+\sigma \widehat{B}_H(T_{\alpha}(t))\}, \label{eq:4} \\
 &&S_0=\widehat{V}(0)>0 \nonumber.
\end{eqnarray}

Making the change of variable, $B_H(t)=\frac{\mu+r_f-r_d}{\sigma}t+\widehat{B}_H(t)$, then we have

\begin{eqnarray}
&&S_t=\widehat{V}(R_\beta(t))=S_0exp\{(r_d-r_f) (T_{\alpha}(t)+\sigma B_H((T_{\alpha}(t))\}, \label{eq:6-1} \\
 &&S_0=\widehat{V}(0)>0.\nonumber
\end{eqnarray}
When the price of the underlying stock $S_t$ satisfies Eq. (\ref{eq:6-1}), we derive an explicit option pricing formula for the European
currency call option. This formula is similar to the Black–Scholes option pricing formula, but with the volatility being different.

We denote the subordinated process $W_{\alpha,H}(t)=B_H(T_{\alpha}(t))$, where $B_H(\tau)$ is a $FBM$ and $T_{\alpha}(t)$ is the inverse $\alpha$-subordinator, which are supposed to be independent. The process $W_{\alpha,H}(t)$ called a subdiffusion process. Particularly, when $H=\frac{1}{2}$, it is a subdiffusion process presented in \cite{magdziarz3,magdziarz4}.

\begin{figure}[H]
  \centering
          \includegraphics[width=1\textwidth]{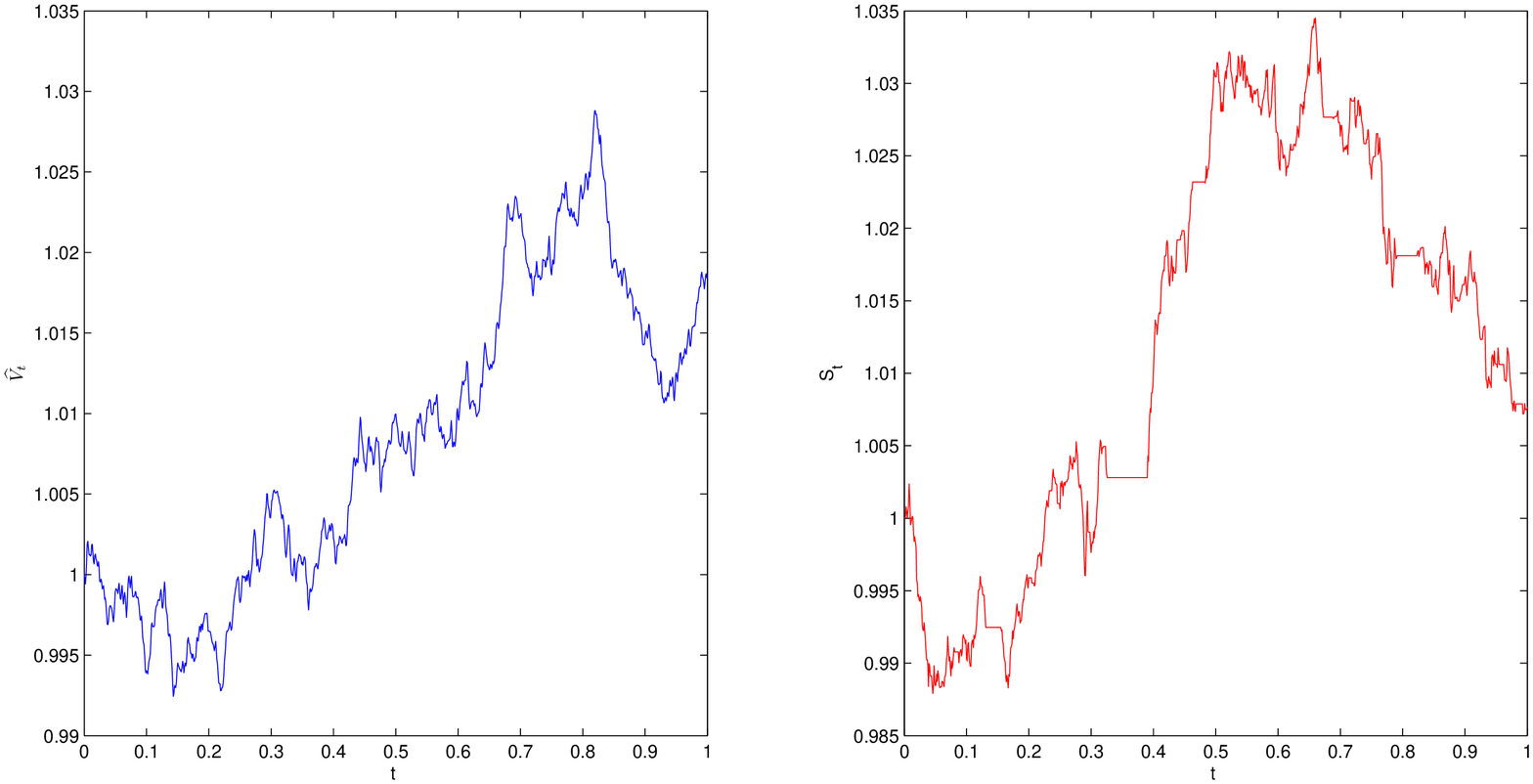}
  \caption{Comparison of the spot exchange rate'sample paths in the $FBM$ model (left) and the subdiffusive $FBM$ model (right) for  $r_d=0.03, r_f=0.02, \alpha=0.9, H=0.8, \sigma=0.1, S_0=1$.}
\label{fig1}
\end{figure}

Fig. \ref{fig1} shows typically the differences and relationships between the sample paths of the spot exchange rate in the $FBM$ model and the subdiffusive $FBM$ model.

The rest of the paper proceeds as follows: In Section \ref{section 3}, we provide an analytic pricing formula for the European currency option in the subdiffusive $FBM$ environment and some Greeks of our pricing model are also obtained. Section \ref{section 4} is devoted to analyze  the impact of scaling and
long-range dependence on currency option pricing. Moreover, the comparison of our subdiffusive $FBM$ model and traditional models is undertaken in this section. Finally, Section \ref{section 5} draws the concluding remarks.

\section{Pricing model for the European call currency option}\label{section 3}
In this section we derive a pricing formula for the European call currency option of the subdiffusive $FBM$
model under the following assumptions:

\begin{enumerate}
\item[(i)] We consider two possible investments: (1) a stock whose price satisfies the equation:
\begin{eqnarray}
S_t=S_0exp\{(r_d-r_f) T_{\alpha}(t)+\sigma  W_{\alpha,H}(t)\},\quad S_0>0,
\label{eq:14}
\end{eqnarray}
where $\alpha\in(\frac{1}{2},1)$,  $H\in[\frac{1}{2},1)$, $2\alpha-\alpha H>1$ and $r_d,$ and, $ r_f$ are the domestic and the foreign interest rates respectively. (2) A money market account:
\begin{eqnarray}
dF_t=r_dF_tdt,
\label{eq:15}
\end{eqnarray}
where $r_d$ shows the domestic interest rate.
\item[(ii)] The stock pays no dividends or other distributions and all securities are perfectly divisible. There are no penalties to
short selling. It is possible to borrow any fraction of the price of a security to buy it or to hold it, at the short-term
interest rate. These are the same valuation policy as in the $BS$ model.

\item[(iii)] There are transaction costs which are proportional to the value of the transaction in the underlying stock. Let k denote
the round trip transaction cost per unit dollar of transaction. Suppose $U$ shares of the underlying stock are bought
$(U>0)$ or sold $(U<0)$ at the price $S_t$, then the transaction cost is given by $\frac{k}{2}|U|S_t$ in either buying or selling. Moreover, trading takes place only at discrete intervals.

\item[(iv)] The option value is replicated by a replicating portfolio $\Pi$ with $U(t)$ units of stock and riskless bonds with value $F(t)$. The value of the option must equal the value of the replicating portfolio to reduce (but not to avoid) arbitrage opportunities
and be consistent with economic equilibrium.

\item[(v)] The expected return for a hedged portfolio is equal to that from an option. The portfolio is revised every $\Delta t$ and hedging
takes place at equidistant time points with rebalancing intervals of (equal) length $\Delta t$, where $\Delta t$ is a finite and fixed,
small time-step.
\begin{rem} From \cite{magdziarz3, guo}, we have  $E(T_{\alpha}^m(t))=\frac{t^{m\alpha}m!}{\Gamma(m\alpha+1)}$. Then, by using $\alpha$-self-similar and non-decreasing sample paths of $T_{\alpha}(t)$, we can obtain that $\alpha$-self-similarand non-decreasing sample paths of $T_{\alpha}(t)$,

\begin{eqnarray}
E\left(\Delta T_{\alpha}(t)\right)=\frac{1}{\Gamma(1+\alpha)}\left[(t+\Delta t)^\alpha-t^\alpha\right]=\frac{t^{\alpha-1}}{\Gamma(\alpha)}\Delta t.
\label{eq:12}
\end{eqnarray}
and
\begin{eqnarray}
E\left((\Delta B_H(T_{\alpha}(t))^2\right)=\left[\frac{t^{\alpha-1}}{\Gamma(\alpha)}\right]^{2H}\Delta t^{2H}.
\label{eq:13}
\end{eqnarray}
\label{re:2-3}
\end{rem}

\end{enumerate}
Let $C=C(t,S_t)$ be the price of a European currency option at time $t$ with a strike price $K$ that matures at time $T$. Then, the pricing formula for currency call option is given by the following theorem

\begin{thm}
$C=C(t,S_t)$ is the value of the European currency call option on the stock $S_t$ satisfied (\ref{eq:6-1}) and the trading takes place
discretely with rebalancing intervals of length $\Delta t$. Then $C$ satisfies the partial differential equation

\begin{eqnarray}
\frac{\partial C}{\partial t}+(r_d-r_f)S_t\frac{\partial C}{\partial S_t}+\frac{1}{2}\widehat{\sigma}^2S_t^2\frac{\partial^2C}{\partial S_t^2}-r_dC=0,
\label{eq:16}
\end{eqnarray}

with boundary condition $C(T,S_T)=\max\{S_T-K,0\}$. The value of the currency call option is

\begin{eqnarray}
C(t,S_t)=S_t e^{-r_f(T-t)}\Phi(d_1)-Ke^{-r_d(T-t)}\Phi(d_2),
\label{eq:17}
\end{eqnarray}

and the value of the put currency option is
\begin{eqnarray}
P(t,S_t)=Ke^{-r_d(T-t)}\Phi(-d_2)-S_t e^{-r_f(T-t)}\Phi(-d_1),
\label{eq:18}
\end{eqnarray}

where
\begin{eqnarray}
d_1&=&\frac{\ln(\frac{S_t}{K})+\left(r_d-r_f\right)(T-t)+\frac{\widehat{\sigma}^2}{2}(T-t)}{\widehat{\sigma}\sqrt{T-t}},\nonumber\\ d_2&=&d_1-\widehat{\sigma}(t)\sqrt{T-t},
\label{eq:19}
\end{eqnarray}

\begin{eqnarray}
\widehat{\sigma}^2=\sigma^2\left[\left(\frac{t^{\alpha-1}}{\Gamma(\alpha)}\right)^{2H}\Delta t^{2H-1}+\sqrt{\frac{2}{\pi}}\frac{k}{\sigma}\left(\frac{t^{\alpha-1}}{\Gamma(\alpha)}\right)^{H}\Delta t^{H-1}\right],
\label{eq:20}
\end{eqnarray}
where $\Phi(.)$ is the cumulative normal distribution function.
\label{th:3-1}
\end{thm}

In what follows, the properties of  the subdiffusive $FBM$ model are discussed, such as Greeks, which summarize how option prices change with respect to underlying variables and are critically important to asset pricing and risk management. The model can be used to rebalance a portfolio to achieve the desired exposure to certain risk. More importantly, by knowing the Greeks, particular exposure can be hedged from adverse changes in the market by using appropriate amounts of other related financial instruments. In contrast to option prices that,
can be observed in the market, Greeks can not be observed and must be calculated given a model assumption. The Greeks are typically computed using a partial differentiation of the price formula.

\begin{thm} The Greeks can be written as follows

\begin{eqnarray}
 \Delta=\frac{\partial C}{\partial S_t}=e^{-r_f(T-t)}\Phi(d_1),
 \label{eq:30}
\end{eqnarray}

\begin{eqnarray}
 \nabla=\frac{\partial C}{\partial K}=-e^{-r_d(T-t)}\Phi(d_2),
 \label{eq:31}
\end{eqnarray}

\begin{eqnarray}
 \rho_{r_d}=\frac{\partial C}{\partial r_d}=K(T-t)e^{-r_d(T-t)}\Phi(d_2),
 \label{eq:32}
\end{eqnarray}

\begin{eqnarray}
 \rho_{r_f}=\frac{\partial C}{\partial r_f}=-S_t(T-t)e^{-r_f(T-t)}\Phi(d_1),
 \label{eq:33}
\end{eqnarray}

\begin{eqnarray}
 \Theta&=&\frac{\partial C}{\partial t}=S_tr_fe^{-r_f(T-t)}\Phi(d_1)-Kr_de^{-r_d(T-t)}\Phi(d_2)\nonumber\\
 &+&S_te^{-r_f(T-t)}\sigma^2(\alpha-1)\frac{t^{\alpha-2}}{\Gamma(\alpha)}\frac{H \big(\frac{t^{\alpha-1}}{\Gamma(\alpha)}\big)^{2H-1}\Delta t^{2H-1}}{\widehat{\sigma}\sqrt{T-t}}(T-t)\Phi'(d_1)\nonumber\\
 &+&S_te^{-r_f(T-t)}\sqrt{\frac{2}{\pi}}k\sigma(\beta-1)\frac{t^{\alpha-2}}{\Gamma(\alpha)}\frac{H \big(\frac{t^{\alpha-1}}{\Gamma(\alpha)}\big)^{H-1}\Delta t^{H-1}}{2\widehat{\sigma}\sqrt{T-t}}(T-t)\Phi'(d_1)\nonumber\\
 &-&S_te^{-r_f(T-t)}\frac{\widehat{\sigma}}{2\sqrt{T-t}}\Phi'(d_1),
 \label{eq:34}
\end{eqnarray}

\begin{eqnarray}
\Gamma=\frac{\partial^2 C}{\partial S_t^2}=e^{-r_f(T-t)}\frac{\Phi'(d_1)}{S_t\widehat{\sigma}\sqrt{T-t}},
\label{eq:35}
\end{eqnarray}

\begin{eqnarray}
\vartheta_{\widehat{\sigma}}=\frac{\partial C}{\partial \widehat{\sigma}}=S_te^{-r_f(T-t)}\sqrt{T-t}\Phi'(d_1).
\label{eq:36}
\end{eqnarray}
\label{th:3-6}
\end{thm}

\begin{rem} The modified volatility without transaction costs $(k=0)$ is given by

 \begin{eqnarray}
\widehat{\sigma}^2=\sigma^2\left[\left(\frac{t^{\alpha-1}}{\Gamma(\alpha)}\right)^{2H}\Delta t^{2H-1}\right],
\label{eq:21}
\end{eqnarray}
 specially if $\alpha\uparrow 1$,
 \begin{eqnarray}
\widehat{\sigma}^2=\sigma^2\Delta t^{2H-1},
\label{eq:23}
\end{eqnarray}
which is consistent with the result in \cite{necula}.

Furthermore, from Eq. (\ref{eq:23}) if $H\uparrow\frac{1}{2}$, then $\widehat{\sigma}^2=\sigma^2$ which is according to the results with the $G-K$ model \cite{garman}.
\label{re:3-3-1}
\end{rem}
Letting $\alpha\uparrow 1$, from Eq. (\ref{eq:20}), we obtain
\begin{rem}
The modified volatility under transaction costs is given by

\begin{eqnarray}
\widehat{\sigma}^2=\sigma^2\left[\Delta t^{2H-1}+\sqrt{\frac{2}{\pi}}\frac{k}{\sigma}\Delta t^{H-1}\right],
\label{eq:22}
\end{eqnarray}
 that is in line with the findings in \cite{wang}.
 \label{re:3-3}
\end{rem}

\section{Empirical Studies}\label{section 4}

The aim of this section is to obtain the minimal price of an option with transaction costs and to show the impact of time scaling $\Delta t$, transaction costs $k$, and subordinator parameter $\alpha$ on the subdiffusive $FBM$ model. Moreover, in the last part we compute the currency option prices using our model and make comparisons with the results of the $G-K$ and $FBM$ models.

Since $\frac{k}{\sigma}<\sqrt{\frac{\pi}{2}}$ often holds (For example: $\sigma=0.1, k=0.01$), from Eq. (\ref{eq:20}) we have
\begin{eqnarray}
\frac{\widehat{\sigma}^2}{\sigma^2}&=&\left(\frac{t^{\alpha-1}}{\Gamma(\alpha)}\right)^{2H}\Delta t^{2H-1}+\sqrt{\frac{2}{\pi}}\frac{k}{\sigma}\left(\frac{t^{\alpha-1}}{\Gamma(\alpha)}\right)^{H}\Delta t^{H-1}\nonumber\\
&\geq&2\left(\frac{t^{\alpha-1}}{\Gamma(\alpha)}\right)^{\frac{3}{2}H}\Delta t^{\frac{3}{2}H-1}\left(\frac{2}{\pi}\right)^{\frac{1}{4}}\left(\frac{k}{\sigma}\right)^{\frac{1}{2}}
,
\label{eq:22-1}
\end{eqnarray}
where $H>\frac{1}{2}$. Then the minimal volatility $\widehat{\sigma}_{min}$ is $\sqrt{2}\sigma\left(\frac{t^{\alpha-1}}{\Gamma(\alpha)}\right)^{\frac{1}{2}}\left(\frac{2}{\pi}\right)^{\frac{1}{2}-\frac{1}{4H}}\left(\frac{k}{\sigma}\right)^{1-\frac{1}{2H}}$
as $\Delta t=\left(\frac{t^{\alpha-1}}{\Gamma(\alpha)}\right)^{-1}\left(\frac{2}{\pi}\right)^{\frac{1}{2H}}\left(\frac{k}{\sigma}\right)^{\frac{1}{H}}$.
Thus the minimal price of an option under transaction costs is represented as $C_{min}(t,S_t)$ with $\widehat{\sigma}_{min}$ in Eq. (\ref{eq:19}).

Moreover, the option rehedging time interval for traders to take is
$\Delta t=\left(\frac{t^{\alpha-1}}{\Gamma(\alpha)}\right)^{-1}\left(\frac{2}{\pi}\right)^{\frac{1}{2H}}\left(\frac{k}{\sigma}\right)^{\frac{1}{H}}$.
The minimal price $C_{min}(t,S_t)$ can be used as the actual price of an option.

\begin{figure}[H]
  \centering
          \includegraphics[width=1\textwidth]{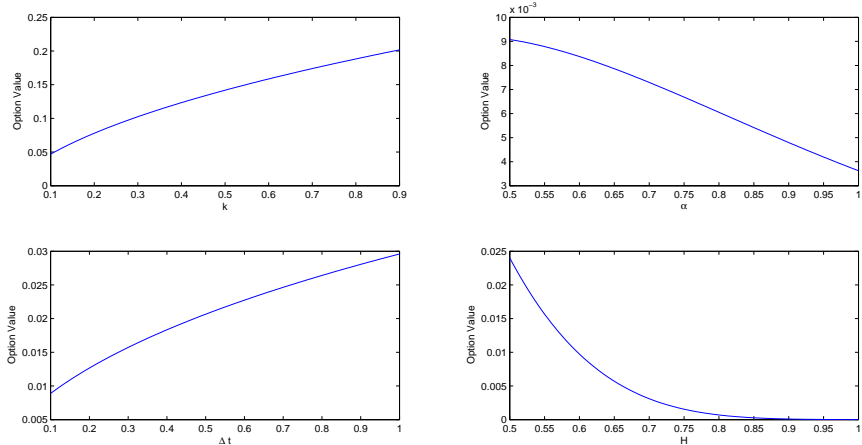}

  \caption{Call currency option values }
  \label{fig4}
\end{figure}

In particular, since $\Delta t <1, \alpha\in(\frac{1}{2},1)$ and $\frac{\partial C}{\partial \widehat{\sigma}}=S_te^{-r_f(T-t)}\frac{\sqrt{T-t}}{\sqrt{2\pi}}e^{-\frac{d^2}{2}}>0,$

\begin{eqnarray}
\frac{\partial\widehat{\sigma}}{\partial H}&=&\sigma\left[2\left(\frac{t^{\alpha-1}}{\Gamma(\alpha)}\right)^{2H}\Delta t^{2H-1}+\sqrt{\frac{2}{\pi}}\frac{k}{\sigma}\left(\frac{t^{\alpha-1}}{\Gamma(\alpha)}\right)^{H}\Delta t^{H-1}\right]\left[\ln\left(\frac{t^{\alpha-1}}{\Gamma(\alpha)}\right)+\ln\Delta t\right]\nonumber\\
&&\times2\left[\left(\frac{t^{\alpha-1}}{\Gamma(\alpha)}\right)^{2H}\Delta t^{2H-1}+\sqrt{\frac{2}{\pi}}\frac{k}{\sigma}\left(\frac{t^{\alpha-1}}{\Gamma(\alpha)}\right)^{H}\Delta t^{H-1}\right]^{-\frac{1}{2}}\nonumber\\
&=&\left[2\left(\frac{t^{\alpha-1}}{\Gamma(\alpha)}\right)^{2H}\Delta t^{2H-1}+\sqrt{\frac{2}{\pi}}\frac{k}{\sigma}\left(\frac{t^{\alpha-1}}{\Gamma(\alpha)}\right)^{H}\Delta t^{H-1}\right]\nonumber\\
&&\times\frac{\sigma^2\left[\ln\left(\frac{t^{\alpha-1}}{\Gamma(\alpha)}\right)+\ln\Delta t\right]}{2\widehat{\sigma}}<0,
\label{eq:22-2}
\end{eqnarray}
and $\frac{\partial C}{\partial H}=\frac{\partial C}{\partial\widehat{\sigma}}\frac{\partial\widehat{\sigma}}{\partial H}$, then we have
\begin{eqnarray}
\frac{\partial C}{\partial H}<0\quad as\, H\in[\frac{1}{2},1),
\label{eq:22-3}
\end{eqnarray}
which displays that an increasing Hurst exponent comes along with a decrease of the option value (See Fig. \ref{fig4}).

On the other hand, if $H\uparrow\frac{1}{2}$, then
\begin{eqnarray}
\widehat{\sigma}_{min}&=& \sqrt{2}\sigma\left(\frac{t^{\alpha-1}}{\Gamma(\alpha)}\right)^{\frac{1}{2}}\left(\frac{2}{\pi}\right)^{\frac{1}{2}-\frac{1}{4H}}\left(\frac{k}{\sigma}\right)^{1-\frac{1}{2H}}\rightarrow\sigma\sqrt{2\left(\frac{t^{\alpha-1}}{\Gamma(\alpha)}\right)}
\label{eq:22-4}
\end{eqnarray}
 and  if $\alpha\uparrow 1$, then $\widehat{\sigma}_{min}\rightarrow\sqrt{2}\sigma$ as $H\uparrow\frac{1}{2}.$

In addition, if $H\uparrow\frac{1}{2}$
\begin{eqnarray}
\Delta t=\left(\frac{t^{\alpha-1}}{\Gamma(\alpha)}\right)^{-1}\left(\frac{2}{\pi}\right)^{\frac{1}{2H}}\left(\frac{k}{\sigma}\right)^{\frac{1}{H}}\rightarrow\left(\frac{t^{\alpha-1}}{\Gamma(\alpha)}\right)^{-1}\left(\frac{2}{\pi}\right)\left(\frac{k}{\sigma}\right)^{2},
\label{eq:22-5}
\end{eqnarray}
and if $\alpha\uparrow 1$, then $\Delta t\rightarrow \left(\frac{2}{\pi}\right)\left(\frac{k}{\sigma}\right)^{2}$ as $H\uparrow\frac{1}{2}.$

Lux and Marchesi \cite{lux1999scaling} have shown that Hurst exponent $H=0.51\pm0.004$ in some cases, so the equations (\ref{eq:22-4}) and (\ref{eq:22-5})
have a practical application in option pricing. For example: if $H\uparrow\frac{1}{2}, \alpha\uparrow 1, k=2\%$ and $\sigma=20\%$, then $\widehat{\sigma}_{min}\rightarrow\frac{\sqrt{2}}{20}$, and $\Delta t\rightarrow\frac{0.02}{\pi}$; and if $H\uparrow\frac{1}{2}, \alpha\uparrow 1, k=0.2\%$ and $\sigma=20\%$, then $\widehat{\sigma}_{min}\rightarrow\frac{\sqrt{2}}{20}$, and $\Delta t\rightarrow\frac{2}{\pi}\times 10^{-4}.$

In the following, we investigate the impact of scaling and long-range dependence on option pricing. It is well known that Mantegna and Stanley \cite{mantegna1995scaling} introduced the method of scaling invariance from the complex science into the economic systems for the first time. Since then, a lot of research for scaling laws in finance has begun. If $H=\frac{1}{2}$ and $k=0$, from Eq. (\ref{eq:20}) we know that $\widehat{\sigma}^2=\sigma^2\left(\frac{t^{\alpha-1}}{\Gamma(\alpha)}\right)$ shows that fractal scaling $\Delta t$ has not any impact on option pricing if a mean self-financing delta-hedging strategy is applied in a discrete time setting, while subordinator parameter $\beta$ has remarkable impact on option pricing in this case. In particular, from Eqs. (\ref{eq:22-4}) and (\ref{eq:22-5}), we know that $\widehat{\sigma}_{min}\rightarrow\sigma\sqrt{2\left(\frac{t^{\alpha-1}}{\Gamma(\alpha)}\right)}$ as $H\approx\frac{1}{2}$ and $\Delta t\rightarrow\left(\frac{t^{\alpha-1}}{\Gamma(\alpha)}\right)^{-1}\left(\frac{2}{\pi}\right)\left(\frac{k}{\sigma}\right)^{2},$ as $H\approx\frac{1}{2}$. Therefore $C_{min}(t,S_t)$ is approximately scaling-free with
respect to the parameter $k$, if $H\approx\frac{1}{2}$,  but is scaling dependent with respect to subordinator parameter $\alpha$. However, $\Delta t\rightarrow\left(\frac{t^{\alpha-1}}{\Gamma(\alpha)}\right)^{-1}\left(\frac{2}{\pi}\right)\left(\frac{k}{\sigma}\right)^{2},$ is scaling-dependent with respect to parameters $k$ and $\alpha$, if $H\approx\frac{1}{2}$. On the other hand, if $H>\frac{1}{2}$ and $k=0$, from Eq. (\ref{eq:21}) we know that $\widehat{\sigma}^2=\sigma^2\left[\left(\frac{t^{\alpha-1}}{\Gamma(\alpha)}\right)^{2H}\Delta t^{2H-1}\right]$, which displays that the fractal
scaling $\Delta t$ and sabordinator parameter $\alpha$ have a significant impact on option pricing. Furthermore, for $k\neq 0$, from Eq. (\ref{eq:19}) we know that option pricing is scaling-dependent in general.

Now, we present the values of currency call option
using subdiffusive $FBM$ model for different parameters. For the sake of simplicity, we will just consider the out-of-the-money case. Indeed, using the same method,
one can also discuss the remaining cases: in-the-money and at-the-money. First, the prices of our subdiffusive $FBM$ model are investigated for some $\Delta t$ and  prices for different exponent parameters. The prices of the call currency option versus its parameters $H, \Delta t, \alpha$ and $k$ are revealed in Fig. \ref{fig4}.  The selected parameters are $S_t=1.4, K=1.5, \sigma=0.1, r_d=0.03, r_f=0.02, T=1, t=0.1,  \Delta t=0.01, k=0.01, H=0.8, \alpha=0.9$. Fig. \ref{fig4} indicates  that, the option price is an increasing function of $k$ and  $\Delta t$, while, it is a decreasing function of $H$ and $ \alpha$.

\begin{figure}
  \centering
          \includegraphics[width=1\textwidth]{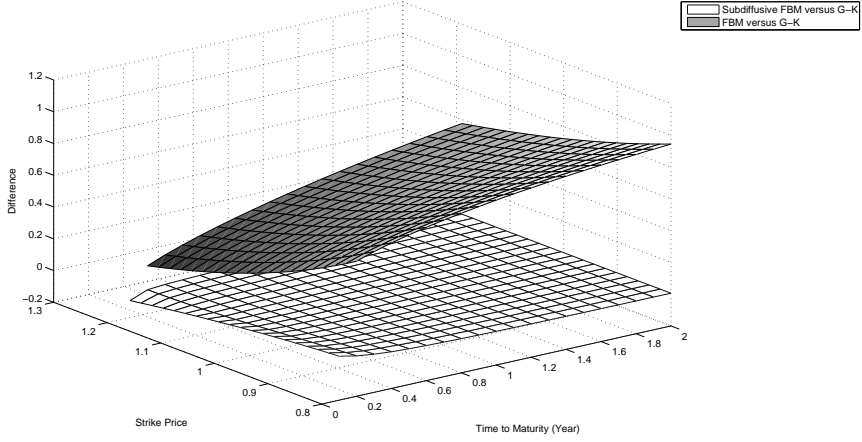}

  \caption{Relative difference between the $G-K$, $FBM$, and subdiffusive $FBM$ models for the in-the-money case}
  \label{fig5}
\end{figure}

\begin{figure}[H]
  \centering
          \includegraphics[width=1\textwidth]{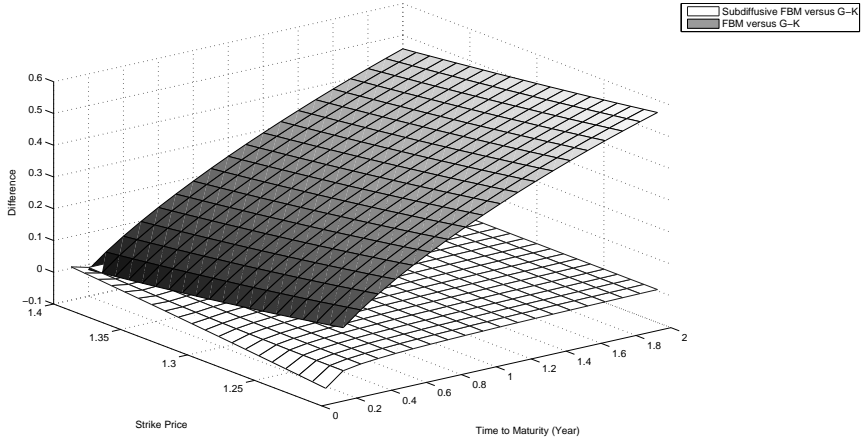}

  \caption{Relative difference between the $G-K$, $FBM$, and subdiffusive $FBM$ models for the out-of-the-money case}
  \label{fig6}
\end{figure}

For a detailed analysis of our model,  the prices  calculated by the $G-K$, $FBM$ and subdiffusive $FBM$ models are compared for both out-of-the-money and in-the-money cases. The following  parameters are chosen: $S_t=1.2, \sigma=0.5, r_d=0.05, r_f=0.01, t=0.1, \Delta t=0.01, k=0.001$, and $ H=0.8$, along with time maturity $T\in [0.1,2]$ , strike price $K\in[0.8,1.19]$ for the in-the-money case and $K\in[1.21,1.4]$ for the out-of-the-money case. Figs. \ref{fig5} and  \ref{fig6} show the theoretical values difference by the $G-K$, $FBM$, and our subdiffusive $FBM$ models for the in-the-money and out-of-the-money, respectively. As indicated in  these figures, the values computed by our subdiffusive $FBM$ model are better fitted to the $G-K$ values than the $FBM$ model for
both in-the-money and out-of-the money cases. Hence, when compared to these figures, our subdiffusive $FBM$ model seems reasonable.

\section{Conclusion}\label{section 5}

Without using the arbitrage argument, in this paper we derive a European
currency option pricing model with transaction costs to capture the
behavior of the spot exchange rate price, where the spot exchange rate follows a subdiffusive $FBM$ with transaction costs.
In discrete time case, we show that the time scaling $\Delta t$ and the Hurst exponent $H$ play
an important role in option pricing with or without transaction costs and option pricing is scaling-dependent. In particular,
the minimal price of an option under transaction costs is obtained.

\section*{Appendix}\label{appendix}

\textbf{Proof of Theorem \ref{th:3-1}.} The movement of $S_t$ on time interval $[t,t+\Delta t)$ of length $\Delta t$ is

\begin{eqnarray}
\Delta S_t&=&S_{t+\Delta t}-S_t=S_t\Big(e^{(r_d-r_f)\Delta T_{\alpha}(t)+\sigma \Delta W_{\alpha,H}(t)}-1\Big)\nonumber\\
&=&S_t\Big((r_d-r_f)\Delta T_{\alpha}(t)+\sigma \Delta W_{\alpha,H}(t)\nonumber\\
&+&\frac{1}{2}((r_d-r_f)\Delta T_{\alpha}(t)+\sigma \Delta W_{\alpha,H}(t))^2\Big)\nonumber\\
&+&\frac{1}{6}S_te^{[\theta((r_d-r_f)\Delta T_{\alpha}(t)+\sigma \Delta W_{\alpha,H}(t))]}\nonumber\\
&\times&\Big((r_d-r_f)\Delta T_{\alpha}(t)+\sigma \Delta W_{\alpha,H}(t)\Big)^3,
\label{eq:43}
\end{eqnarray}

here  $\theta=\theta(t,\Delta t)\in(0,1)$ is a random variable according to process $S_t$.

Note that
\begin{eqnarray}
& &\frac{1}{6}S_te^{[\theta((r_d-r_f)\Delta T_{\alpha}(t)+\sigma \Delta W_{\alpha,H}(t))]}\leq\frac{1}{6}S_te^{(r_d-r_f)\Delta T_{\alpha}(T)}.e^{\sigma|\Delta W_{\alpha,H}(t)|}\nonumber\\
&\leq&\frac{1}{6}S_te^{(r_d-r_f) T_{\alpha}(T)}.e^{\sigma| W_{\alpha,H}(t)|}.e^{\sigma| W_{\alpha,H}(t+\Delta t)|}.
\label{eq:44}
\end{eqnarray}

and for $m\in N$,

\begin{eqnarray}
E\big(e^{m(r_d-r_f) T_{\alpha}(T)}\big)&=&\sum_{j=0}^\infty\frac{(m(r_d-r_f))^j}{j!}E\big(T_{\alpha}(T)^j\big)\nonumber\\
&=&\sum_{j=0}^\infty\frac{(m(r_d-r_f) T^\alpha)^j}{\Gamma(j\alpha+1)}\nonumber\\
&=&E_\alpha(m(r_d-r_f) T^\alpha)<+\infty,
\label{eq:45}
\end{eqnarray}

where $E_\alpha(.)$ is the Mittage-Lefller function \cite{samko}.

Based on Lemmas 2.1 and 2.2 in \cite{gu} and Eq. (\ref{eq:45}), we have

\begin{eqnarray}
\Delta t^{2\varepsilon}.\frac{1}{6}S_te^{\theta((r_d-r_f)\Delta T_{\alpha}(t)+\sigma\Delta W_{\alpha,H}(t))}=o(\Delta t^\varepsilon),
\label{eq:46}
\end{eqnarray}

\begin{eqnarray}
& &\frac{1}{6}S_te^{\theta((r_d-r_f)\Delta T_{\alpha}(t)+\sigma \Delta W_{\alpha,H}(t))}((r_d-r_f)\Delta T_{\alpha}(t)+\sigma \Delta W_{\alpha,H}(t))^3\nonumber \\&&
=\Delta t^{-2\varepsilon}.o(\Delta t^\varepsilon).\big(o(\Delta t^{\alpha-\varepsilon})+o(\Delta t^{\alpha H-\varepsilon})\big)^3\nonumber\\&&
=o(\Delta t^{3\alpha H-4\varepsilon})=o(\Delta t).
\label{eq:47}
\end{eqnarray}

Then,

\begin{eqnarray}
\Delta S_t&=&(r_d-r_f) S_t\Delta T_{\alpha}(t)+\sigma S_t\Delta W_{\alpha,H}(t)\nonumber\\
&+&\frac{1}{2}\sigma^2S_t(\Delta W_{\alpha,H}(t))^2+o(\Delta t).
\label{eq:48}
\end{eqnarray}

By using the Taylor expansion we get

\begin{eqnarray}
\Delta C(t,S_t)&=&\frac{\partial C}{\partial t}\Delta t+\frac{\partial C}{\partial S_t}\Delta S_t+\frac{1}{2}\frac{\partial^2 C}{\partial S_t^2}\Delta S_t^2 \nonumber\\&&
+\frac{1}{2}\frac{\partial^2 C}{\partial t^2}\Delta t^2+\frac{\partial^2 C}{\partial S_t\partial t}\Delta t\Delta S_t+o(\Delta t^{3\alpha H-\varepsilon})\nonumber\\&&
=\frac{\partial C}{\partial t}\Delta t+\frac{\partial C}{\partial S_t}\Delta S_t+\frac{1}{2}\frac{\partial^2 C}{\partial S_t^2}\Delta S_t^2+o(\Delta t)\nonumber\\&&
=\frac{\partial C}{\partial t}\Delta t+(r_d-r_f) S_t\frac{\partial C}{\partial S_t}\Delta T_{\alpha}(t)+\sigma S_t\frac{\partial C}{\partial S_t}\Delta W_{\alpha,H}(t)\nonumber\\&&
+\frac{1}{2}\sigma^2 S_t\frac{\partial C}{\partial S_t}(\Delta W_{\alpha,H}(t))^2\nonumber\\&&
+\frac{1}{2}\sigma^2 S_t^2\frac{\partial^2 C}{\partial S_t^2}(\Delta W_{\alpha,H}(t))^2+o(\Delta t).
\label{eq:49}
\end{eqnarray}

From Eq. (\ref{eq:45}), we obtain that $\frac{\partial^2 C}{\partial S_t^2}$, $\frac{\partial^3 C}{\partial S_t^3}$, $\frac{\partial^2 C}{\partial C\partial t}$ is $o(\Delta t^{\frac{1}{2}(1-H\alpha)^{-\varepsilon}})$ and

\begin{eqnarray}
\Delta\big(\frac{\partial C}{\partial S_t}\big)=\frac{\partial^2 C}{\partial S_t\partial t}\Delta t+\frac{\partial^2 C}{\partial S_t^2}\Delta S_t+\frac{1}{2}\frac{\partial^3 C}{\partial S_t^3}\Delta S_t^2+o(\Delta t),
\label{eq:50}
\end{eqnarray}
and
\begin{eqnarray}
\big|\Delta\big(\frac{\partial C}{\partial S_t}\big)\big|.S_{t+\Delta t}= \sigma S_t^2\big|\frac{\partial^2 C}{\partial S_t^2}\big||\Delta W_{\alpha,H}(t)|+o(\Delta t).
\label{eq:51}
\end{eqnarray}

Moreover, from assumptions (iii) and (iv), it is found that the change in the value of portfolio $\Pi_t$ is
\begin{eqnarray}
\Delta \Pi_t&=&U_t\big(\Delta S_t+r_fS_t\Delta t\big)+\Delta F_t-\frac{k}{2}|\Delta U_t|S_{t+\Delta t}\nonumber\\
&=&U_t\big(\Delta S_t+r_fS_t\Delta t\big)+r_dF_t\Delta t\nonumber\\
&-&\frac{k}{2}|\Delta U_t|S_{t+\Delta t}+o(\Delta t),
\label{eq:52}
\end{eqnarray}
where the number of bonds $U_t$ is constant during time-step $ \Delta t$. From assumption (v), $C(t,S_t)$ is replicated by portfolio $\Pi(t)$. Thus, at time points $\Delta t$, $2\Delta t$, $3\Delta t,... , $ we have  $C(t,S_t)=U_tS_t+F_t$ and $F_t=\frac{\partial C}{\partial S_t}$. Therefore, according to Eqs. (\ref{eq:48})-(\ref{eq:52}) we have

\begin{eqnarray}
\Delta \Pi&=&\frac{\partial C}{\partial S_t}\big[(r_d-r_f) S_t\Delta T_{\alpha}(t)+\sigma S_t \Delta W_{\alpha,H}(t)+\frac{1}{2}\sigma^2 S_t(\Delta W_{\alpha,H}(t))^2+r_fS_t\Delta t\big]\nonumber\\
&+&r_dF_t\Delta t-\frac{k}{2}\big|\Delta\big(\frac{\partial C}{\partial S_t}\big)\big|.S_{t+\Delta t}+o(\Delta t)  \nonumber\\
&=&\frac{\partial C}{\partial S_t}\big[(r_d-r_f)  S_t\Delta T_{\alpha}(t)+\sigma S_t \Delta W_{\alpha,H}(t)+\frac{1}{2}\sigma^2 S_t(\Delta W_{\alpha,H}(t))^2+r_fS_t\Delta t\big]\nonumber\\
&+&\big(C(t,S_t)-S_t\frac{\partial C}{\partial S_t}\big)r_d\Delta t-\frac{k}{2} \sigma S_t^2\big|\frac{\partial^2C}{\partial S_t^2}\big||\Delta W_{\alpha,H}(t)|+o(\Delta t).
\label{eq:53}
\end{eqnarray}
Consequently,
\begin{eqnarray}
\Delta \Pi-\Delta C&=&\big(r_dC-(r_d-r_f)S_t\frac{\partial C}{\partial S_t}-\frac{\partial C}{\partial t}\big)\Delta t-\frac{1}{2}\sigma^2 S_t^2\frac{\partial^2C}{\partial S_t^2}(\Delta W_{\alpha,H}(t))^2\nonumber \\
&-&\frac{k}{2}\sigma S_t^2\big|\frac{\partial^2C}{\partial S_t^2}\big||\Delta W_{\alpha,H}(t)|+o(\Delta t).
\label{eq:54}
\end{eqnarray}

The time subscript, $t$ has been suppressed. As expected, using  Eq. (\ref{eq:54}), (iv), Remark \ref{re:2-3}, and
 \cite{leland} we infer
\begin{eqnarray}
E(\Delta \Pi-\Delta C)&=&\big(r_dC-(r_d-r_f)S_t\frac{\partial C}{\partial S_t}-\frac{\partial C}{\partial t}\big)\Delta t\nonumber\\
&-&\frac{1}{2}\big[\frac{t^{\alpha-1}}{\Gamma(\alpha)}\big]^{2H}\Delta t^{2H}\sigma^2S_t^2\frac{\partial^2C}{\partial S_t^2}-\frac{1}{2}\sqrt{\frac{2}{\pi}}k \sigma S_t^2\big[\frac{t^{\alpha-1}}{\Gamma(\alpha)}\big]^{H}\Delta t^{H}\big| \frac{\partial^2C}{\partial S_t^2}\big|\nonumber\\
&=&\Big(r_dC-(r_d-r_f)S_t\frac{\partial C}{\partial S_t}-\frac{\partial C}{\partial t}
-\frac{1}{2}\big[\frac{t^{\alpha-1}}{\Gamma(\alpha)}\big]^{2H}\Delta t^{2H-1}\sigma^2S_t^2\frac{\partial^2C}{\partial S_t^2}\nonumber \\
&-&\frac{1}{2}\sqrt{\frac{2}{\pi}}k \sigma S_t^2\big[\frac{t^{\alpha-1}}{\Gamma(\alpha)}\big]^{H}\Delta t^{H-1}\big| \frac{\partial^2C}{\partial S_t^2}\big|\Big)\Delta t=0.
\label{eq:55}
\end{eqnarray}

Thus, from Eq. (\ref{eq:55}) we  can derive
\begin{eqnarray}
r_dC\nonumber&=&(r_d-r_f)S_t\frac{\partial C}{\partial S_t}+\frac{\partial C}{\partial t}
+\frac{1}{2}\big[\frac{t^{\alpha-1}}{\Gamma(\alpha)}\big]^{2H}\Delta t^{2H-1}\sigma^2S_t^2\frac{\partial^2C}{\partial S_t^2}\nonumber \\ &&
+\frac{1}{2}\sqrt{\frac{2}{\pi}}k \sigma S_t^2\big[\frac{t^{\alpha-1}}{\Gamma(\alpha)}\big]^{H}\Delta t^{H-1}\big| \frac{\partial^2C}{\partial S_t^2}\big|.
\label{eq:56}
\end{eqnarray}

We define $\widehat{\sigma}^2(t)$ as follows
\begin{eqnarray}
\widehat{\sigma}^2=\sigma^2\Big(\big[\frac{t^{\alpha-1}}{\Gamma(\alpha)}\big]^{2H}\Delta t^{2H-1}+\sqrt{\frac{2}{\pi}}k \sigma^{-1} \big[\frac{t^{\alpha-1}}{\Gamma(\alpha)}\big]^{H}\Delta t^{H-1}\Big).
\label{eq:57}
\end{eqnarray}
where $\frac{\partial^2C}{\partial S_t^2}$ is ever positive for the ordinary European currency call option without transaction costs, if  the same conduct of $\frac{\partial^2C}{\partial S_t^2}$ is postulated here and $\widehat{\sigma}(t)$ remains fixed during the time-step $[t, \Delta t)$. Then, from Eqs. (\ref{eq:56}) and (\ref{eq:57}) we obtain

\begin{eqnarray}
\frac{\partial C}{\partial t}+(r_d-r_f)S_t\frac{\partial C}{\partial S_t}+\frac{1}{2}\widehat{\sigma}^2S_t^2\frac{\partial^2 C}{\partial S_t^2}-r_dC=0.
\label{eq:58}
\end{eqnarray}

Followed by

\begin{eqnarray}
C=C(t,S_t)=S_te^{-r_f(T-t)} \Phi(d_1)-Ke^{-r_d(T-t)}\Phi(d_2),
\label{eq:59}
\end{eqnarray}
and
\begin{eqnarray}
d_1&=&\frac{\ln(\frac{S_t}{K})+\big(r_d-r_f\big)(T-t)+\frac{\widehat{\sigma}^2}{2}(T-t)}{\widehat{\sigma}\sqrt{T-t}},\nonumber\\
 d_2&=&d_1-\widehat{\sigma}\sqrt{T-t}.
\label{eq:60}
\end{eqnarray}

\textbf{Proof of Theorem \ref{th:3-6}.} First, we derive a general formula. Let $y$ be one of the influence factors. Thus

\begin{eqnarray}
\frac{\partial C}{\partial y}&=&\frac{\partial
S_te^{-(r_f)(T-t)}}{\partial
y}\Phi(d_1)+S_te^{-r_f(T-t)}\frac{\partial
\Phi(d_1)}{\partial y}\nonumber\\
&-&\frac{\partial Ke^{-r_d(T-t)}}{\partial
y}\Phi(d_2)-Ke^{-r_d(T-t)}\frac{\partial \Phi(d_2)}{\partial y}
\label{eq:61}
\end{eqnarray}

But

\begin{eqnarray}
\frac{\partial \Phi(d_2)}{\partial y}&=&\Phi'(d_2)\frac{\partial
d_2}{\partial y}\nonumber\\
&=&\frac{1}{\sqrt{2\pi}}e^{-\frac{d_2^2}{2}}\frac{\partial
d_2}{\partial y}\nonumber\\
&=&\frac{1}{\sqrt{2\pi}}\exp\big(-\frac{(d_1-\widehat{\sigma}\sqrt{T-t})^2}{2}\big)\frac{\partial
d_2}{\partial y}\nonumber\\
&=&\frac{1}{\sqrt{2\pi}}e^{-\frac{d_1^2}{2}}\exp\big(d_1\widehat{\sigma}\sqrt{T-t)}\big)\exp\big(-\frac{\widehat{\sigma}^2(T-t)}{2}\big)\frac{\partial
d_2}{\partial y}\nonumber\\
&=&\frac{1}{\sqrt{2\pi}}e^{-\frac{d_1^2}{2}}\exp\big(\ln
\frac{S_t}{K}+(r_d-r_f)(T-t)\big)\frac{\partial
d_2}{\partial y}\nonumber\\
&=&\frac{1}{\sqrt{2\pi}}e^{-\frac{d_1^2}{2}}\frac{S}{K}
\exp\big((r_d-r_f)(T-t)\big)\frac{\partial
d_2}{\partial y}.
\label{eq:62}
\end{eqnarray}

Then
\begin{eqnarray}
\frac{\partial C}{\partial y}&=&\frac{\partial S_te^{-(r_f)(T-t)}}{\partial
y}\Phi(d_1)-\frac{\partial Ke^{-r_d(T-t)}}{\partial
y}\Phi(d_2)\nonumber\\
&+&S_te^{-r_f(T-t)}\Phi'(d_1)\frac{\partial\widehat{\sigma}\sqrt{T-t)}}{\partial y}.
\label{eq:63-1}
\end{eqnarray}
Substituting in (\ref{eq:63-1}) we get the desired Greeks.


\bibliographystyle{elsarticle-num}
\bibliography{../../reference}
\end{document}